\documentclass[amssymb, reprint, showpacs, aip, floatfix]{revtex4-1}

\makeatletter
\let\origsection\subsubsection
\renewcommand\subsubsection{\@ifstar{\subsubstarsection}{\nostarsubsubsection}}
\newcommand\nostarsubsubsection[1]{\subsubsectionprelude\origsection{#1}\subsubsectionpostlude}

\newcommand\subsubsectionprelude{\vspace{0em} }
\newcommand\subsubsectionpostlude{\vspace{-1em} }
\makeatother

\usepackage{graphicx}			
\usepackage{dcolumn}			
\usepackage{bm}					
\usepackage[colorlinks=true,linkcolor=blue,citecolor=blue,urlcolor=blue]{hyperref}
\usepackage{float}			
\usepackage{amsmath,amsfonts}	

\usepackage{setspace}		

\usepackage{lineno}   			
\usepackage{ragged2e}
\usepackage{subcaption}
\DeclareCaptionJustification{justified}{\justifying}
\usepackage{multirow}

\usepackage{units}

\def\ie{{\em i.e., }}
\def\eg{{\em e.g., }}
\def\etc{{\em etc.}}
\def\SignalTrain{{\tt SignalTrain}}

\usepackage[dvipsnames]{xcolor}

\begin{document}

\title{SignalTrain: Profiling Audio Compressors with Deep Neural Networks}
\author{Scott H. Hawley}
\email{scott.hawley@belmont.edu}
\affiliation{Department of Chemistry \& Physics,  Belmont University, Nashville, TN USA}
\author{Benjamin Colburn}
\affiliation{ARiA Acoustics, Washington, DC USA}

\author{Stylianos I. Mimilakis}
\affiliation{Fraunhofer Institute for Digital Media Technology, Ilmenau, Germany}

\date{\today}

\begin{abstract}
In this work we present a data-driven approach for predicting the behavior of (\ie profiling) a given non-linear audio signal processing effect (henceforth ``audio effect").
Our objective is to learn a mapping function that maps the unprocessed
audio to the processed by the audio effect to be profiled, using time-domain samples.
To that aim, we employ a deep auto-encoder model that is conditioned on both time-domain samples and the control parameters of the target audio effect.
As a test-case study, we focus on the offline profiling of two dynamic range compression audio effects,
one software-based and the other analog.
Compressors were chosen because they are a widely used and important
set of effects and because their parameterized nonlinear time-dependent nature makes them a challenging problem for a system aiming to profile ``general'' audio effects.
Results from our experimental procedure show that the primary
functional and auditory characteristics of the compressors can be
captured, however there is still sufficient audible noise
to merit further investigation before such methods are applied
to real-world audio processing workflows.
\end{abstract}

\maketitle

\section{\label{sec:1} Introduction}
The ability to digitally model musical instruments and audio effects allows for multiple desirable properties,\citep{smith_physical_2011} among which are
i) portability -- virtual instruments and software effects require no space or weight;
ii) flexibility -- many such effects can be stored and accessed together and quickly modified;
iii) signal to noise -- often can be higher with digital effects;
iv) centralized, automated control;
v) repeatability -- digital effects can be exactly the same, as opposed to physical systems which may require calibration; and
vi) extension -- the development of digital effects involves fewer constraints than their real-world counterparts.

The process of constructing such models
has traditionally been performed using one of two main approaches. One approach is the physical
simulation of the processes involved,\citep{smith_physical_2011} whether
these be acoustical processes such as reverberation\citep{valimaki_more_than_fifty} or ``virtual analog modeling'' of circuit elements.\citep{yeh_numerical_2008, pakarinen_recent_2011,eichas2017virtual}  The other main approach has been to emulate the requisite audio features via signal processing
techniques which seek to capture the salient aspects of the sounds and transformations under consideration.
Both of these approaches are typically performed with the goal of faithfully reproducing one particular effect, such as audio compressors.\cite{stikvoort1986,kroning2011,floru1999attack,gerat2017virtual,simmer2006}

Rather than modeling one particular effect, a different class of systems are those which
can
`profile' and `learn' to mimic the tonal effects of other units. One popular commercial example is the Kemper Profiler Amplifier,\citep{kemper} which can learn to emulate the sounds of amplifiers and speaker cabinets in the user's possession, to enable them to store and easily transport a virtual array of analog gear.  Another product in this category is the ``ToneMatch'' feature of Fractal Audio's Axe-Fx,\citep{axefx} which supplies a large number of automatically-tunable pre-made effects units including reverberation, delay, equalization, and formant processing.

The present paper involves efforts toward the goal of profiling `general' audio effects.  For systems which are linear and time-invariant (LTI), one can develop finite impulse response (FIR) filters, \eg for convolution reverb effects.  But for systems which involve nonlinearity and/or time-dependence, more sophisticated approaches are required.

Deep learning has demonstrated great utility at such diverse audio signal processing tasks as   classification,\citep{choi2017convolutional, fonseca2017acoustic} onset detection,\citep{schluter_onset}
source separation,\citep{paris_sourcesep} event detection,\citep{stowell2015detection}
dereverberation,\citep{wang_dereverb,arifianto2018dereverberation}  denoising,\citep{rethage2018wavenet} formant
estimation,\citep{dissen2019formant} remixing,\citep{pons2016remixing} and synthesis,\citep{engel_neural_2017, tacotron, char2wav, wavegan} as well
as dynamic range compression to automate the
mastering process.\citep{mimilakis2016deep}
In the area of audio component modeling, deep learning
has been used to model tube amplifiers\citep{damskagg_tube}
and most recently guitar distortion pedals.\citep{distortion2}
Besides creating specific effects, efforts have been underway to explore how varied are the types of effects which can be learned from a single model,\citep{martinez_modeling_2018} to which this paper comprises a contribution.

A challenging goal in deep learning audio processing is to devise models that operate directly on the raw waveform signals, in the time domain, known as ``end-to-end" models.\citep{dieleman2014end}
Given that the raw waveform data exists in the time domain, there are questions as to whether an end-to-end formulation is most suitable,\citep{francesc_timedomain} however it has been shown to be useful nevertheless.
Our approach is end-to-end, however, we make use of a spectral representation
within the autoencoders and for regularization.

Our efforts in this array have been focused on modeling dynamic range compressors for three reasons:
1. They constitute a `desirable' problem to solve: Many audio production practitioners
rely on the specific operational characteristics of certain compression units,
and as such compression modeling is a key area of interest for practical
application.
2. They constitute a `hard' problem to solve: As noted earlier, existing
methods are sufficient to implement a variety of linear and/or time-independent
effects.  Our own deep learning investigations (unpublished) demonstrated that
effects such as echo or distortion could be modeled via Long Short-Term Memory (LSTM) cells, but compressors proved to be `unlearnable' to our networks.
This present work therefore describes one solution to this problem.
3. It is our contention that compressors represent a set of capabilities
that would be required for modeling more general effects.

In this study we are not estimating compressor parameters,\citep{simmer2006}
although deep neural networks have recently shown proficiency
at this task as well.\citep{sheng2019_siamese}  Rather, being given parameters
associated with input-output pairs of audio data, we
synthesize audio by means of a network which learns
to emulate given mappings subject to these parameters.
The hope is that by
performing well on the challenging problem of
dynamic range
compression, such a network could also prove useful for learning
other audio effects as well.

It is a common occurrence for researchers in machine learning to associate memorable nomenclature to denote their models.\citep{tacotron, van_den_oord_wavenet_2016}  Given that the goal our system is successively
approximate the audio signal chain through a process of training, we refer to the computer code as \SignalTrain.\footnote{Source code and datasets
accompanying this paper will be publicly released on GitHub.com pending acceptance for publication.}

This paper proceeds as follows: In Section \ref{sec:design}, we describe the problem specification, the design of the neural network architecture, its training procedure, and the dataset. In Section \ref{sec:results} we relate results for two compressor models, one digital and one analog.
Finally we offer some conclusions in section \ref{sec:concl} and outline some avenues for future work.

\section{\label{sec:design} System Design}

\subsection{Problem Specification}

\subsubsection{Overview}

The objective is to accurately model the input-output characteristics of a wide range of musical signal processing
effects, and their parameterized controls, in a {\em model-agnostic} manner.  That is to say, not to merely
infer certain control parameters which are then used in
conjunction with pre-made internal effect modules (\eg
as is done by Axe-FX.)\citep{axefx} We apply our method to
the case of compressors in this paper, but we operate no
internal compressor model -- the system {\em learns
what a compressor is} in the course of training using a
large variety of training signals and control settings.

We conceive of the task as a supervised learning regression problem, performed in an end-to-end manner.  While other approaches have made use of techniques such as $\mu$-law companding and one-hot encoding to formulate the task as a classification problem,\citep{van_den_oord_wavenet_2016} we have not yet done so.  Rather than predicting one audio sample (\ie time-series value) at a time, we map a range of inputs to a range of outputs, \ie we window the audio into ``windows.''  This allows for both speed in computation as well as the potential for modeling non-causal behavior such as reverse-audio effects or time-alignment.  \footnote{The system could be modified to predict one sample at a time, however our experience with this model has found this practice to be neither necessary nor helpful.}

 \vspace{-10pt}
\subsubsection{Compressor Effects Used}
We trained against two software compressors, with similar controls but different time scales.
The effect we designate ``Comp-4C'' which operates
in a sequential manner
(later samples explicitly depend on earlier samples) and has four
controls for Threshold, Ratio, Attack and Release.
The other formulation, ``Comp-4C-Large,'' allows for wider ranges of the control parameters.
For an analog effect we used a Universal Audio LA-2A, output audio for a wide range of input
audio as we varied the Peak Reduction knob and the Comp/Lim switch. (The input and output gain
knobs were left fixed in the creation of the dataset.)
These effects are summarized in
Table \ref{tab:comp_effects}.

\begin{table}[h]
    \begin{tabular}{l|l|p{3.6cm}}
     Effect Name & Type & Controls: Ranges \\
        \hline \hline
      Comp-4C& Software &Threshold: -30--0\,dB\newline Ratio: 1--5\newline Attack: 1--40\,ms\newline Release: 1--40\,ms \\
       \hline
     Comp-4C-Large& Software & Threshold:{-50}--0\,dB\newline Ratio: 1.5--10\newline Attack: 1--1000\,ms\newline Release: 1--1000\,ms \\
      \hline
     LA-2A & Analog & Comp/Lim Switch: 0/1\newline Peak Reduction: 0--100\\
     \hline
    \end{tabular}
    \caption{Compressor effects trained. Comp-4C and Comp-4C-Large allow different control ranges but use the same Python code, which is available in supplementary materials.\citep{suppl_materials}
    The physical LA-2A unit also has controls for input gain and
    output gain, but these were not varied for this study.
    (All control ranges are rescaled
    to \mbox{[-0.5,0.5]} for input to the neural network.)
    Other compressor features such as ``knee," side-chaining, multi-band compression, {\em etc.}, were not included in this study.}
    \label{tab:comp_effects}
\end{table}

 \vspace{-20pt}
\subsubsection{Data Specification and Error Estimates}
\label{sec:causality}
Typically audio effects are applied to an entire ``stream'' of data from beginning
to end, yet it is not uncommon for digital audio processors to be presented with
only a smaller ``window'' (also referred to as a ``chunk,''
``frame'', ``input buffer,'' \etc)
of the most recent audio, of a duration usually determined by computational
requirements such as memory and/or latency.
For time-dependent effects such as
compressors, the size of the window can have repercussions as information
preceding the window boundary will necessarily propagate into the window currently
under consideration. This introduces a concern over ``causality,'' occurring over a timescale
given by the exponential decay due to the compressor's attack and release controls.

This suggests two different ways to approach training, and two different ways to
specify the dataset of pairs
of input audio and target output audio.  The first we refer to as ``streamed target'' (ST)
data, which is the usual method of applying the audio effect to the entire stream at
once.  The second we refer to ``windowed target'' (WT) data, in which the effect
is applied sequentially to individual windows of input.
WT data will necessarily
contain transient errors (compared to ST data) occurring on a frequency of the
inverse of the window duration.  If however one
adds a ``lookback buffer,'' i.e. making the length
of the output shorter than that of the input,
then this ``lookback'' can be chosen to
be large enough that transient errors in the WT data  decay (exponentially) below
the ``noise floor'' before the output is generated.
The goal of this study is to produce ST data as accurately as possible, as it corresponds
to the normal application of audio effects, but WT data is in some sense ``easier''
to learn.  Indeed, in our early attempts with the LA-2A compressor and ST data, the model was
not able to learn {\em at all}, because the lookback buffer was not long enough.

\begin{figure}[ht]
  {
  \begin{subfigure}[b]{\columnwidth}
  \centering
     \includegraphics[width=0.491\linewidth]{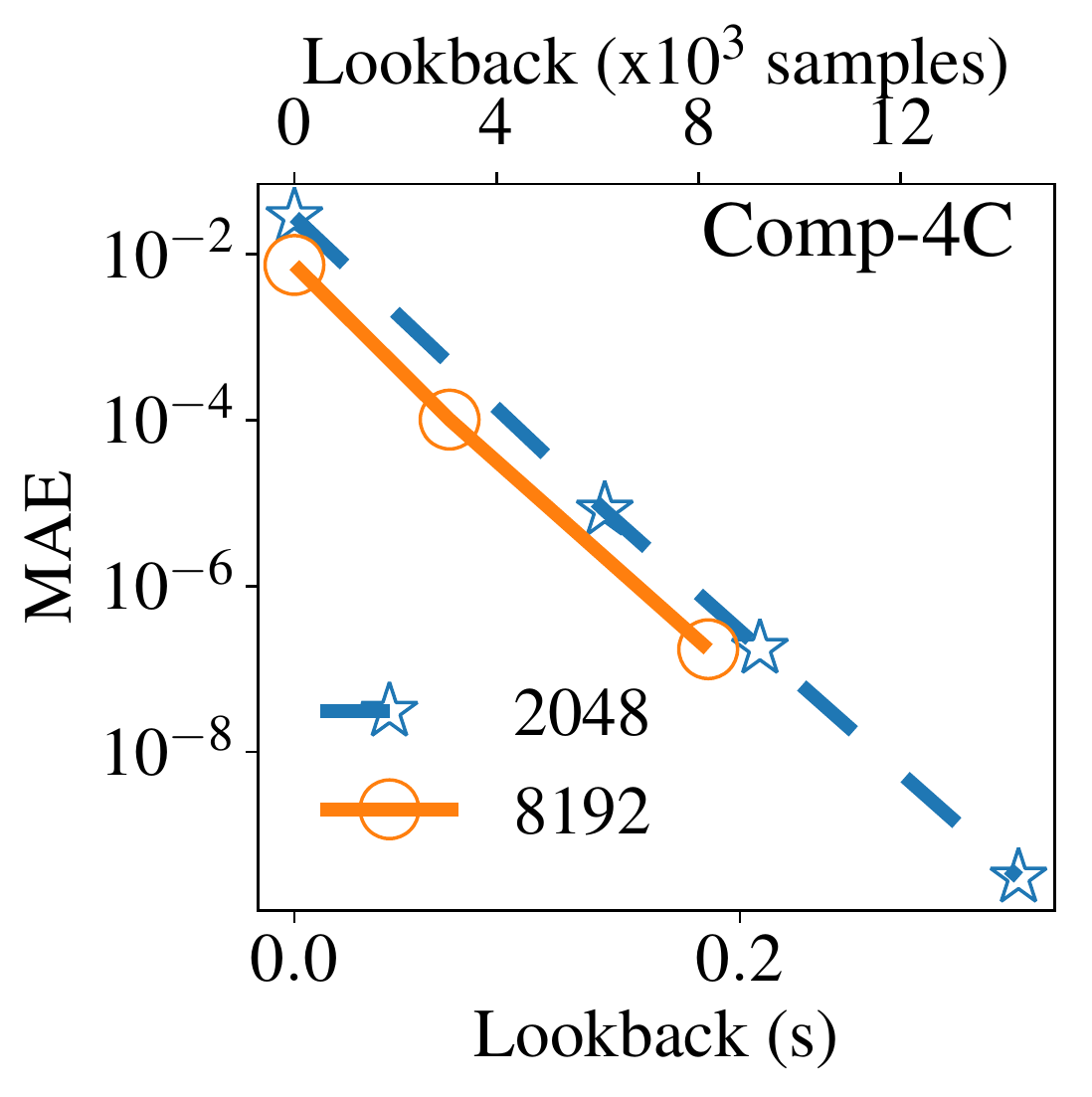}
      \includegraphics[width=0.491\linewidth]{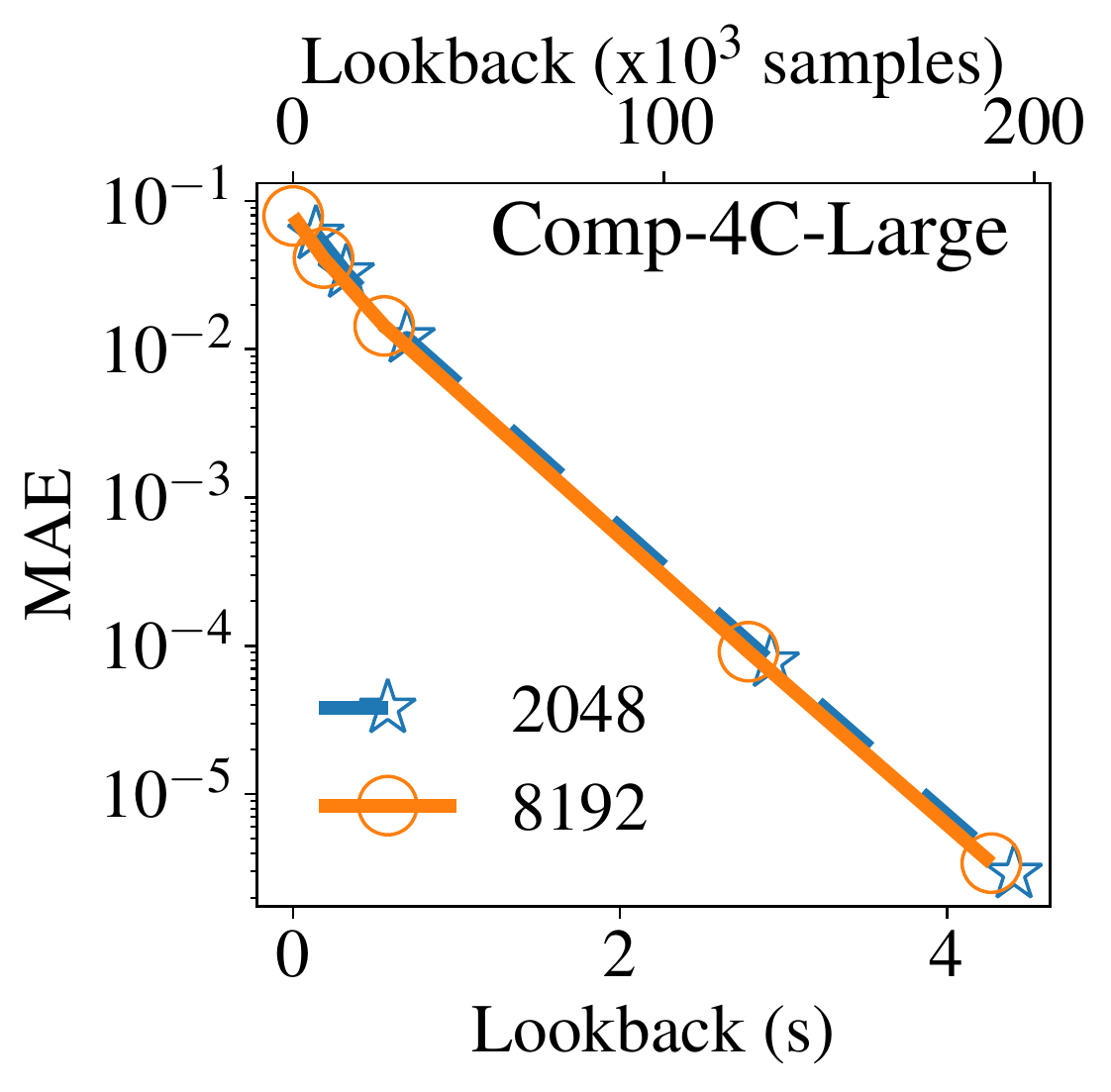}
  \end{subfigure}
  }
\vspace{-5mm}
\caption{\label{fig:lookback}{
Mean Absolute Error (MAE) between streamed target (ST) data vs. windowed target (WT) data, for the Comp-4C compressor effect as a function of lookback buffer size at \mbox{rm 44.1 kHz}.  This represents a theoretical limit for the accuracy of the neural network model, \ie assuming perfect modeling by the neural network, ``causality'' considerations imply that it cannot achieve an error lower than that of the Comp-4C effect itself operating in a WT manner.
This provides a way to estimate the recommended (minimum) size of input and lookback
buffers to use when training the model for a given error goal.
}}
\end{figure}

The difference between ST and WT data constitutes a lower bound on the error
produced by our neural network model: we do
not expect the model to perform better than the ``true'' effect applied to WT data.
The dependence of this error bound on the size of the lookback buffer
can be estimated in a straightforward way,
and can provide guidance on the size of buffer that should be used when
training the model.  Such estimates are shown in Figure \ref{fig:lookback}.   In order to allow for low enough error while
not putting too great a strain on computational resources, we
will choose model sizes with lookback windows sufficient to allow a lower bound on the error in the range of $10^{-5}$ to $10^{-4}$.

\subsection{Model Specification}

\subsubsection{Network architecture.}
The architecture of the \SignalTrain{}
model consists of the front-end, the autoencoder-like module, and the back-end module. The proposed architecture shares some similarities with the U-Net\cite{unet} and TFNet\cite{tfnet} architectures. In more details, the front-end module is comprised by
a set of two 1-D convolution operators that are responsible for producing a signal sub-space similar to a time-frequency decomposition, yielding magnitude and phase features. The autoencoder module consists of two deep neural networks for processing individually the magnitude and phase information of the front-end module. Each deep neural network in this autoencoder consists of 7 fully connected, feed-forward neural networks (FC). It should be denoted that the ``bottleneck'' latent space of each deep neural network is additionally conditioned on the control variables of the audio effect module that are represented as one-hot encoded vectors.

\begin{figure}[ht]
\includegraphics[width=0.9\columnwidth]{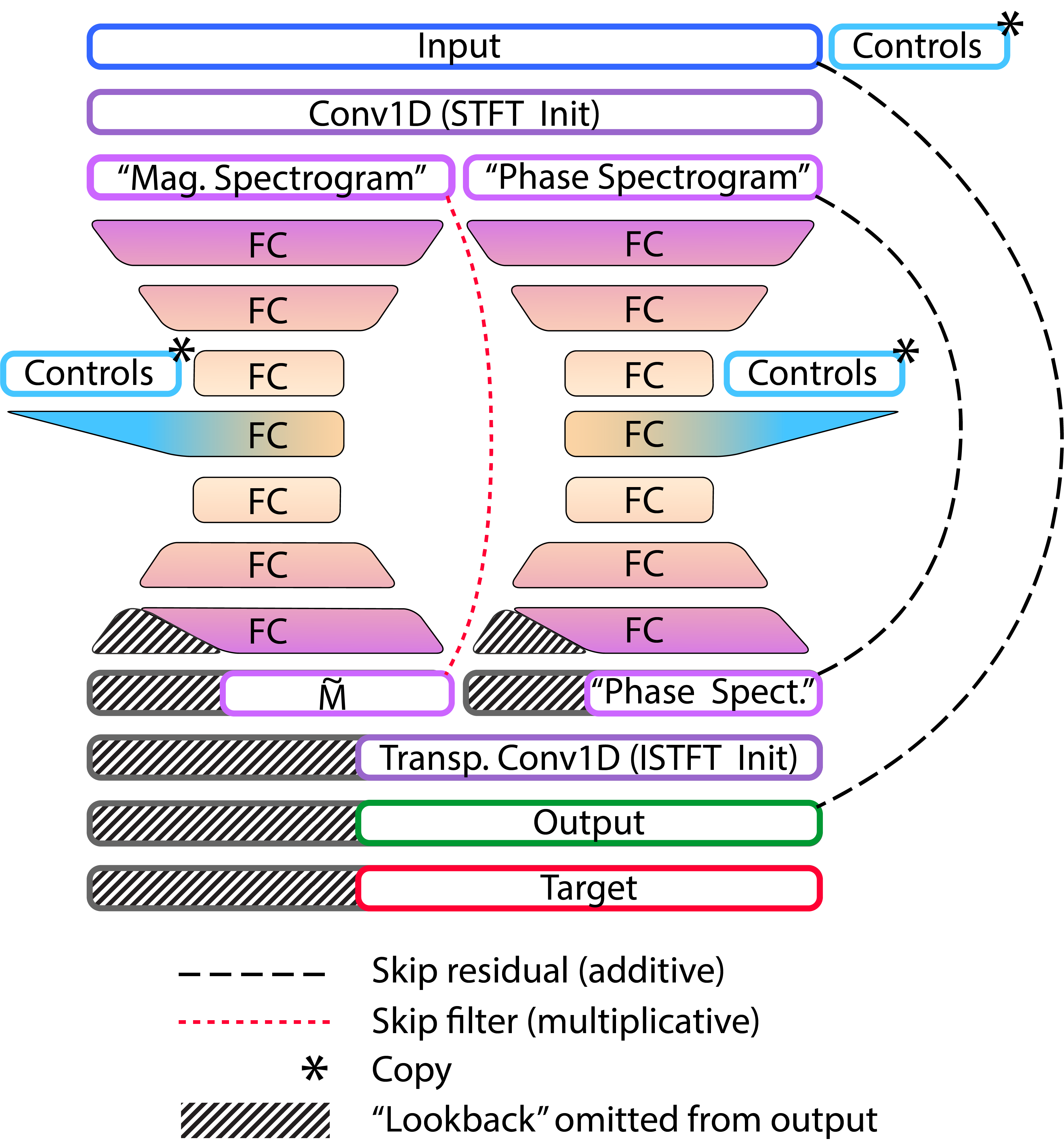}
\caption{\label{fig:1}{
Diagram of the \SignalTrain{} network architecture.
The designations `STFT,' `ISTFT,' `Magnitude' and `Phase' are used for simplicity, since the complex-valued discrete Fourier transform is used to initialize the weights of the 1-D convolution layers. Except for  1-D convolution layers at the beginning and end, all layers are fully-connected (FC) -- \ie ``linear'' (or ``dense'') layers -- with ELU\cite{elu} activations.  This is a notable difference from other models which use convolutions for all layers, \eg U-Net.\citep{unet}
Typically our output and target
waveforms are smaller than the input, \eg coinciding with the last $\nicefrac{1}{4}$ of the input.  The difference in size between input and
output (indicated by the cross-hatched region designated
``lookback'') means the autoencoder is `asymmetric.'
On the autoencoder output, the ``magnitude spectrogram
designated $\tilde{M}$ is used for regularizing the loss, Eq. (\ref{eq:loss}).
}}
\vspace{-10pt}
\end{figure}

Figure \ref{fig:1} illustrates the neural network architecture for the \SignalTrain{} model, which essentially learns a mapping function from the un-processed to the processed audio,
by the audio effect to be profiled, and is conditioned on the vector of the effect's controls  (\eg the ``knobs'').
In order to obtain the predicted output waveform, the back-end module uses another set of two 1-D transposed convolutional operators. Similarly to the analysis front-end, the initialization of the back-end is using the bases of the discrete Fourier transform. It should be stated that all the weights are subject to optimization and are expected to vary during the training of the model. The frame and hop sizes used
for the convolutions are 1024 and 384 samples, respectively.

Unlike some other proposed architectures which use convolutional layers,\cite{unet} we use fully-connected (FC) layers that have shared-weights with respect to the sub-space dimensionality (\ie the frequencies). That is done for two reasons. The first reason is that the number of the parameters inside the model is dramatically reduced, and secondly we preserve the location of the magnitude and phase information of the original signal. Essentially, the operations carried by each deep neural network in the autoencoder module can be seen as non-linear affine transformations of the transpose time-frequency patches (spectrograms) of the input signal.
Furthermore, we apply residual (additive) skip connections\citep{resnet} inspired by U-Net\citep{unet}
and ``skip filter'' (multiplicative) connection for the magnitude only.\cite{mimilakis2016deep}  These skip connections dramatically improve the speed of training, and can be viewed in three complementary ways: allowing information to propagate further through the network,
smoothing the loss surface,\citep{loss_landscape_and_resids} and/or allowing the network
to compute formidable perturbations subject to the goal of profiling an audio effect.

In the middle of the autoencoder, we concatenate values of the effect controls (\eg threshold, ratio, \etc) and ``merge'' these via an additional FC layer. The first layer
of the autoencoder maps the number of time frames in the spectrograms to 64, with
subsequent layers shrinking (or, on the output side, growing) this by factors of 2.
The resulting model has approximately 4 million trainable parameters.

\subsection{\label{subsec:2:3} Training Procedure}

\subsubsection{Loss function.}
\begin{figure}[ht]
\includegraphics[width=3.4in]{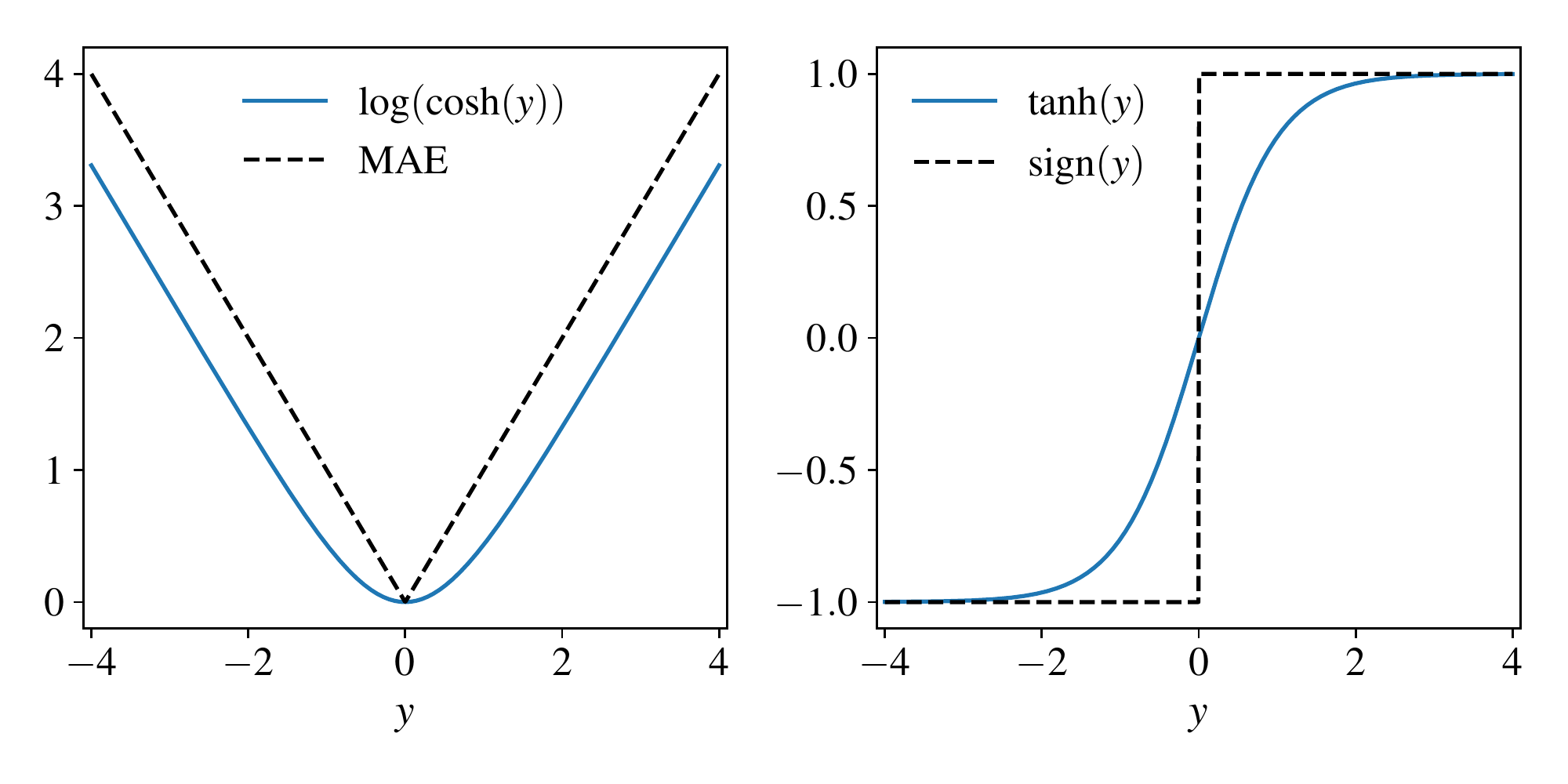}
\vspace{-8mm}
\caption{\label{fig:loss_and_lr}{
Left: The log-cosh loss function and its comparison to the more commonly known MAE (\ie, L1 norm divided by number of elements).
Right: The gradients of log-cosh and MAE are $\tanh(x)$ and ${\rm sgn}(x)$, respectively.  The presence of a continuous derivative for log-cosh allows for superior training via gradient descent compared to MAE.
}}
\end{figure}

We use a log-cosh loss function,\citep{chen2019logcosh}
which has similar properties to
the MAE (\ie L1 norm divided by the number of elements) in the sense that
it forces the predicted signal to closely follow the target signal
at all times, however the roundness of the log-cosh function near zero allows for significantly better training at low loss values than does L1, which has a discontinuous derivative at zero.
This is illustrated in Figure \ref{fig:loss_and_lr}.

Furthermore we include an L1 regularization term with a small
coefficient $\lambda$ (\eg 2e-5), consisting of the magnitude spectrogram $\tilde{M}$ from the output side of the autoencoder, weighted exponentially by frequency-bin number $f_k$ to help reduce high-frequency noise in the predicted output.  Thus the equation for the loss function is given by
\begin{equation}
    \label{eq:loss}
{\rm Loss} = \log\left[\cosh{\left(\tilde{y}-y\right)}\right] + \lambda\exp[(f_k)\,^{\alpha}]\cdot |\tilde{M}|_{L1},
\end{equation}
where $\tilde{y}$ and $y$ are the predicted and target outputs, respectively, and $\alpha=1$ implies exponential weighting by frequency bin $f_k$, and $\alpha=0$ means no such
weighting.

\vspace{-20pt}
\subsubsection{Training Data.}
\begin{figure}[ht]
\includegraphics[width=3.4in]{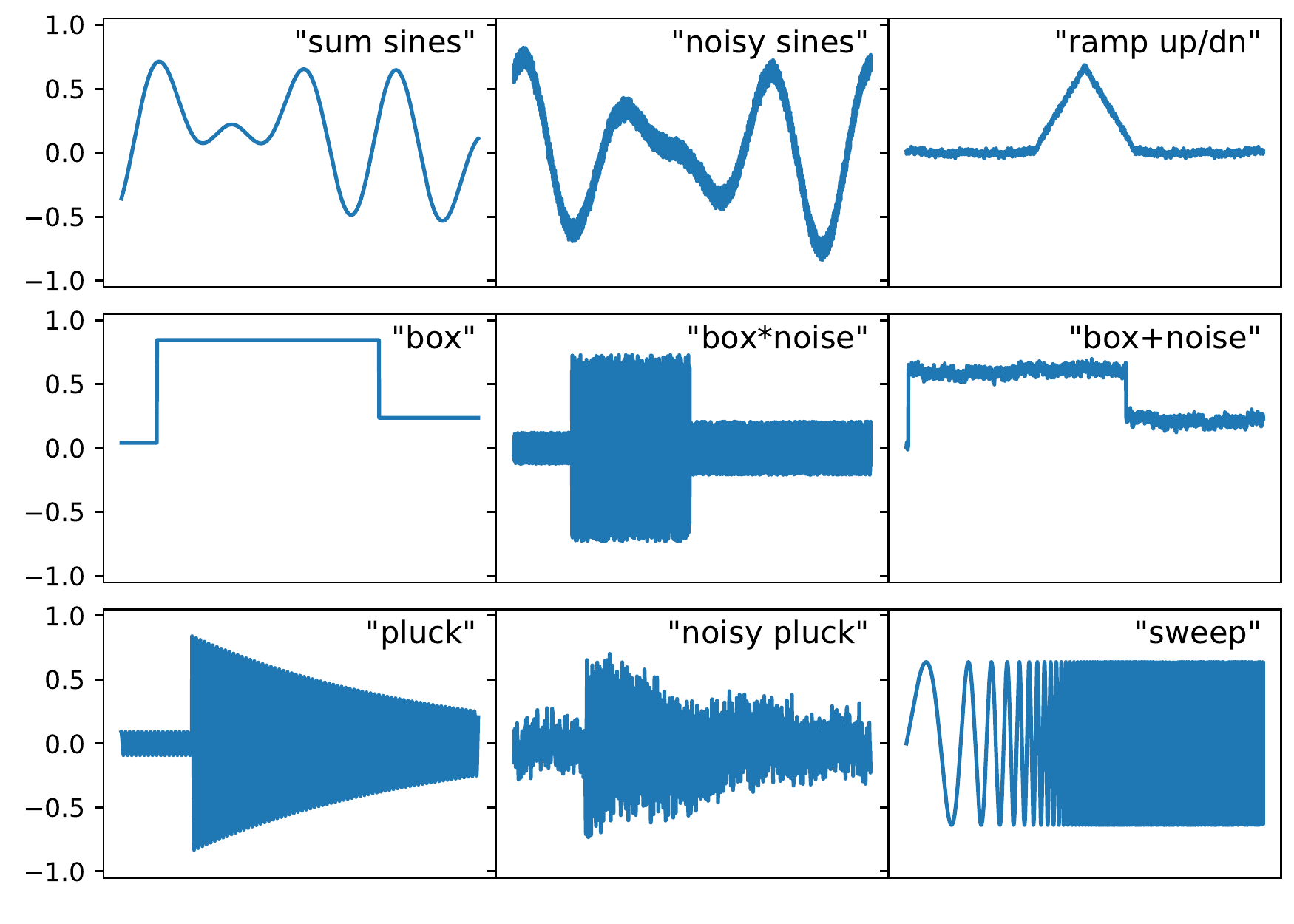}
\vspace{-8mm}
\caption{\label{fig:synth_sounds}{
Example waveforms of randomly synthesized sounds used in addition to music recordings.
All relevant expects of these sounds,
(\eg starting and ending amplitudes, frequencies, decay rates, onsets, cutoffs,
slopes, noise levels, phases, \etc) vary randomly throughout the dataset.  For software
effects such as Comp-4C, such data can be generated ``on the fly'' indefinitely,
whereas for the LA-2A we used a fixed number of such synthetic sounds.
}}
\end{figure}

Simply training on a large amount of musical audio files is
not necessarily the most efficient way to train the network -- depending on
the type of effect being profiled, some signals may be more `instructive' than others.
A compressor requires numerous transients of significant size, whereas
an echo (or `delay') effect may train most efficiently on uncorrelated
input signals (\eg white or pink noise).  Therefore, we augment a dataset of music
recordings with randomly-generated sounds intended to provide both dynamic
range variation and broadband frequency coverage.  Examples of these are shown in Figure \ref{fig:synth_sounds}.

By virtue of the automation afforded by software effects such as Comp-4C, we can
train {\em indefinitely} using randomly-synthesized signals which change during each
iteration.
But for the LA-2A, we created an large (20 GB) input dataset of public
domain musical sounds and randomly-generated test sounds, concatenated these
and divided the result into (unique) files of 15-minute duration, using
a fixed increment of ``5" on the LA-2A's Peak Reduction knob between recordings, for both settings of the
Comp/Lim switch.
For prerecorded (\ie non-synthesized) audio, windows from the input (and for ST data, target) data are copied from random locations in the audio files, along with the control settings used. Data augmentation is
applied only in the form of randomly flipping the phase of inputs and targets.

To achieve the results in this paper, we trained for two days (see ``Implementation,'' below) on what corresponded to approximately 2000 hours of audio sampled at 44.1 kHz (or ~130 GB if it were stored on disk).  As a performance metric, we keep a separate, fixed ``validation set'' of  approximately 12 minutes of audio; all results displayed in this paper are for validation data, \ie on data which the network has not ``seen'' before.

The arrangement of this data is ``maximally shuffled,'' \ie we find that training is significantly more smooth and stable when {\em the control settings are randomly changed for each data window within each mini-batch.}  Trying to train using the same knob settings for large sequences of inputs -- as one might expect to do by taking a lengthy pair of (input-output) audio clips obtained at one effect setting and breaking them up into a sequential mini-batch of training windows --  results in unstable training in the sense that the (training and/or validation) loss varies much more erratically and, overall, decreases much more slowly than for the `maximally shuffled' case in which one varies the knob settings with every window.  This shuffling from window to window is
another reason why our model is not autoregressive: because we wish to learn to model the controls with the effect.

 \vspace{-10pt}
\subsubsection{Initialization.}
When starting from scratch, weights are initialized
randomly except for the weights connecting to the input and output layers, which are
initialized according to a Discrete Fourier Transform (DFT), and its inverse transform, respectively. These are subsequently allowed to evolve as training proceeds.

 \vspace{-10pt}
\subsubsection{Optimizer.}
For the different task of image
classification on the ImageNet dataset,\citep{imagenet_cvpr09} the combination of Adam\citep{adam} with weight decay\citep{weight_decay}
 has been shown\citep{adamw_superconv_2018} to be among the
fastest training methods available when combined with learning rate scheduling.  We also adopt this combination for our problem.

 \vspace{-10pt}
\subsubsection{Learning Rate Scheduling.}
\label{sec:lrsched}
An important feature, found to decrease both final
validation loss values and the number of epochs required to reach them, is the use of learning rate
scheduling, \ie adjusting the value of the
learning rate dynamically during the course of gradient-based optimization, rather than keeping the learning rate static.
We follow the ``1-cycle'' policy,\citep{smith1} which incorporates cosine annealing,\citep{cosine_annealing} in the manner
popularized by the Fasti.ai team.\cite{gugger}
Compared to using a static learning rate, the 1-cycle policy allowed
us to reach roughly 1/10th the error in 1/5 the time.

 \vspace{-10pt}
\subsection{Implementation}
\vspace{-10pt}
The \SignalTrain{} code was written in Python using the PyTorch\citep{pytorch}
library along with Numba for speeding up certain subroutines.
Development and training was primarily conducted on a desktop computer with two NVIDIA Titan X GPUs.  Late the project we upgraded to two RTX 2080Ti GPUs, which, with the benefit of NVIDIA's ``Apex'' mixed-precision (MP) training library,\footnote{https://github.com/NVIDIA/apex} yielded speedup of 1.8x over the earlier runs.
MP training involves computing losses at full
precision while performing inference using ``half-precision'' (\ie FP16) representations,
which have a machine epsilon on the order of $1e-3$, and thus applications of MP are more commonly associated with classification tasks than regression tasks.  While our results obtained from MP and those from full-precision calculations for our
regression task are not strictly identical, they show
no significant differences, even at loss values on the order of $1e-5$.
For the GPUs used, we found by experimentation that a mini-batch size of 200 offered the best training performance
per wall-clock execution time, and each ``epoch'' consisted of 1000 batches of randomly-sampled windows from the audio files (or synthesized on the fly).

\section{\label{sec:results} Results}

\subsection{Software Compressor: ``Comp-4C''}
\vspace{-10pt}
We ported MATLAB code to Python for a single-band, hard-knee compressor with four
controls: threshold, ratio, attack and release times.\cite{hackaudio}
(This compressor implements no side-chaining, make-up gain or other features.)
As it is a software compressor, the training data could be generated ``on the
fly," choosing control (`knob') settings randomly according to some probability
distribution (\eg uniform, or a beta distribution to emphasize the endpoints
\footnote{Initially we chose control settings  according to a symmetric beta
distribution, with the idea that by mildly emphasizing the endpoints of the
control ranges, the model would learn more efficiently, however experience
showed no  performance enhancement compared to choosing from a uniform distribution.}).
This synthesis allows for a virtually limitless size of the training dataset.
Our early experiments used such a dataset, but given that intended goal
of this system is to profile systems within a finite amount of time, and
particularly {\em analog} effects which would typically require the creation
of a finite set of recordings, we chose to emulate the intended use case for
analog gear, namely a finite dataset in which the control knob settings are
equally spaced, with 10 settings per control.

\begin{figure}[b]
\includegraphics[width=\columnwidth]{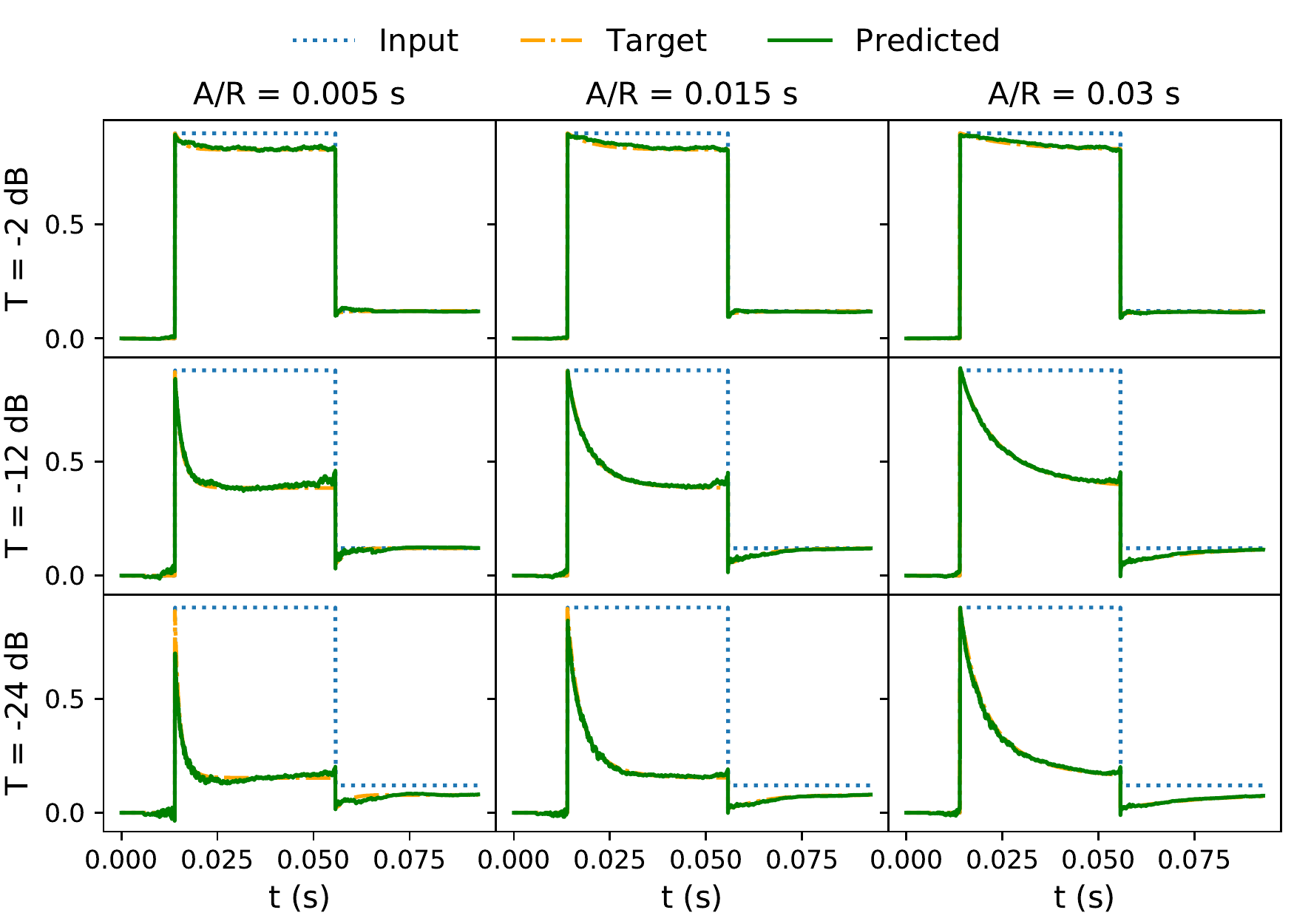}
\caption{\label{fig:boxcollage}{Sample model step-response performance for the Comp-4C effect using WT data on a domain of 4096 samples at 44.1 kHz, for various values of threshold (T) and attack-release (A/R, set equivalently). In all graphs, the ratio=3. See Figure \ref{fig:boxcollage_diff} for a plot of the difference between predicted and target outputs, and Supplemental Materials \cite{suppl_materials} for audio samples and an interactive
demo with various input waveforms and adjustable parameters.
}}
\end{figure}

\begin{figure}[ht]
\includegraphics[width=\columnwidth]{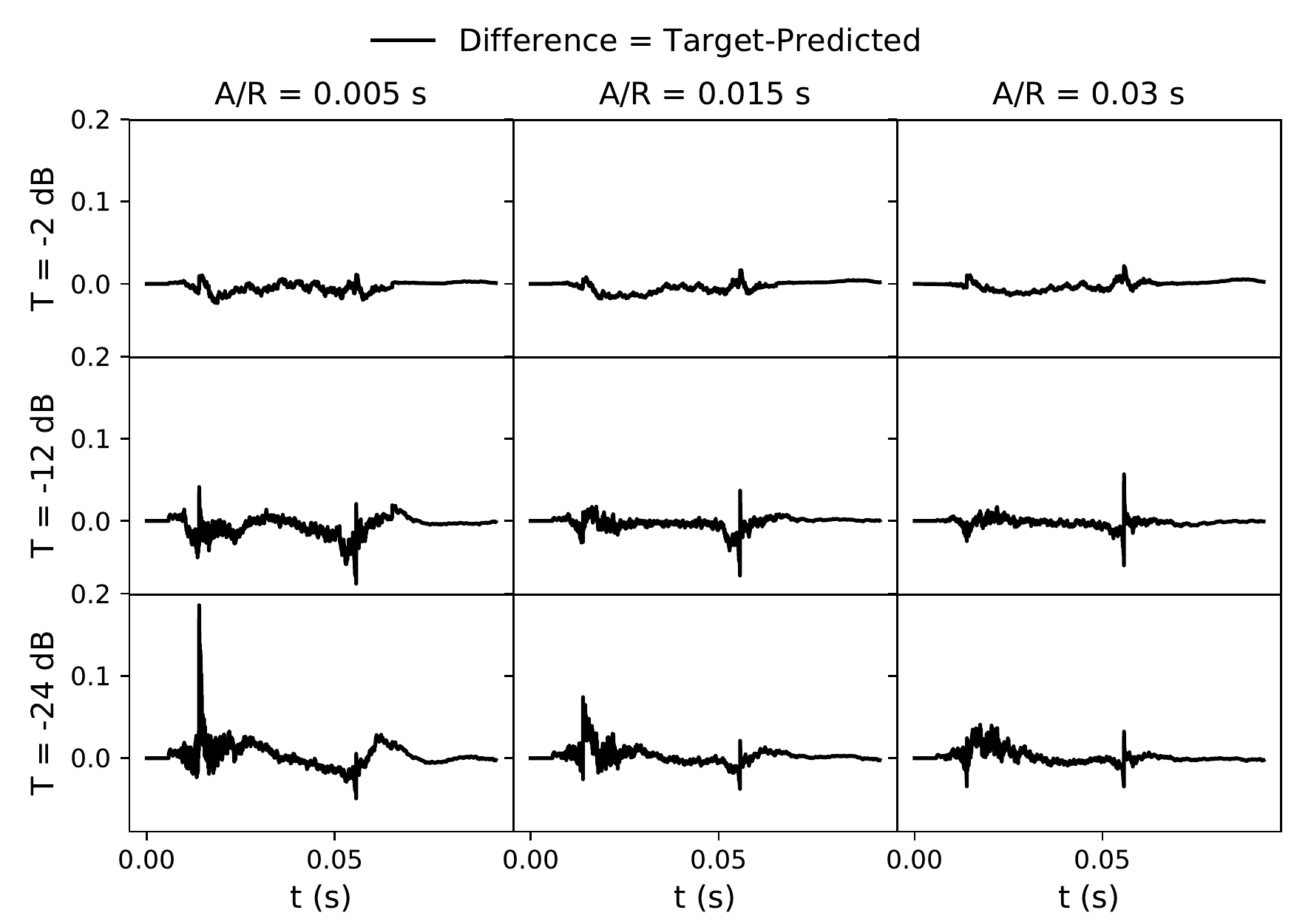}
\caption{\label{fig:boxcollage_diff}{
The difference between predicted and target outputs for the step responses shown in Figure \ref{fig:boxcollage}.  We see the largest errors occur precisely at the step discontinuities,
likely due to inadequate approximation in the ``spectral'' representation within the model.
}}
\vspace{-5pt}
\end{figure}

Figure \ref{fig:boxcollage} shows the performance of the model compared to
the target audio, for the case of a step-response, a common diagnostic
signal for compressor performance.\citep{simmer2006,giannoulis2012, eichas2017virtual}
The predicted values follow the target closely enough that we show their
differences in Figure \ref{fig:boxcollage_diff}.  Key differences occur
at the discontinuities themselves (especially at low attack times),
and we see that the predictions tend to ``overshoot'' slightly on at the
release discontinuity (likely due to slight errors in the phase in the spectral
decomposition in the model), but that in between and after the discontinuities
the predictions and target match closely.

\begin{figure}[ht]
\includegraphics[width=\columnwidth]{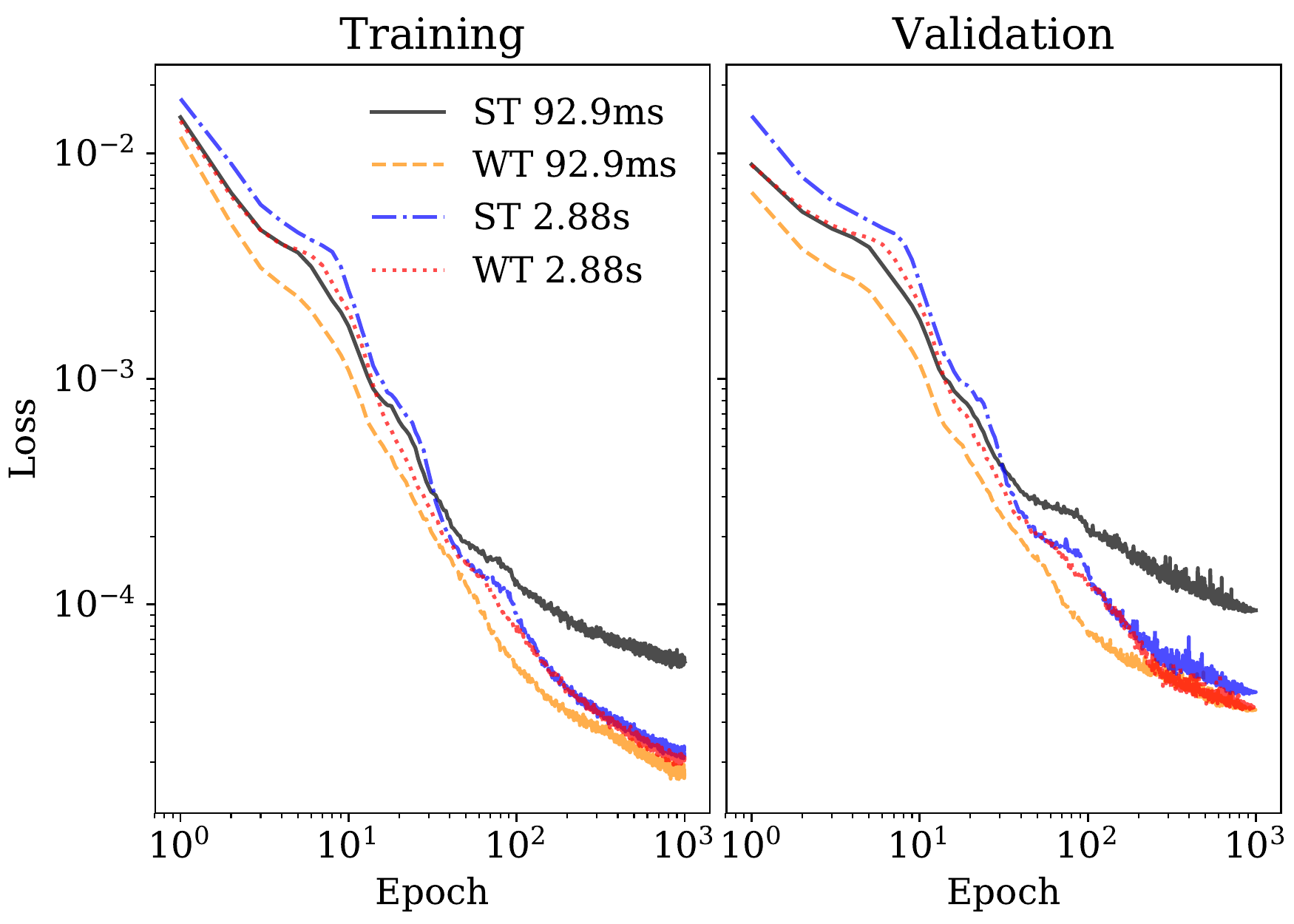}
\caption{\label{fig:comp4c_train}{Typical loss on Training \& Validation sets for Comp-4C-Large effect, while training for ST and WT data, for two different lookback window lengths.
(Because our data is randomly-sampled, ``Epoch'' does not refer to a complete pass through
the dataset, but rather the arbitrary selection of 1000 mini-batches.)
}}
\end{figure}

As noted in Section \ref{sec:causality}, the size of the lookback
window can have an effect on the error bounds.  Figure \ref{fig:comp4c_train} shows that the loss on the Validation set
to be consistent with estimates obtained for the
cases depicted in Figure \ref{fig:lookback}.
And yet listening to these examples (see Supplemental Materials\citep{suppl_materials}) one notices noise in the predicted
results, suggesting that the lookback window size (or ``causality
noise'') is not the only source of error in the model.

\begin{figure}[h]
\includegraphics[width=0.8\columnwidth]{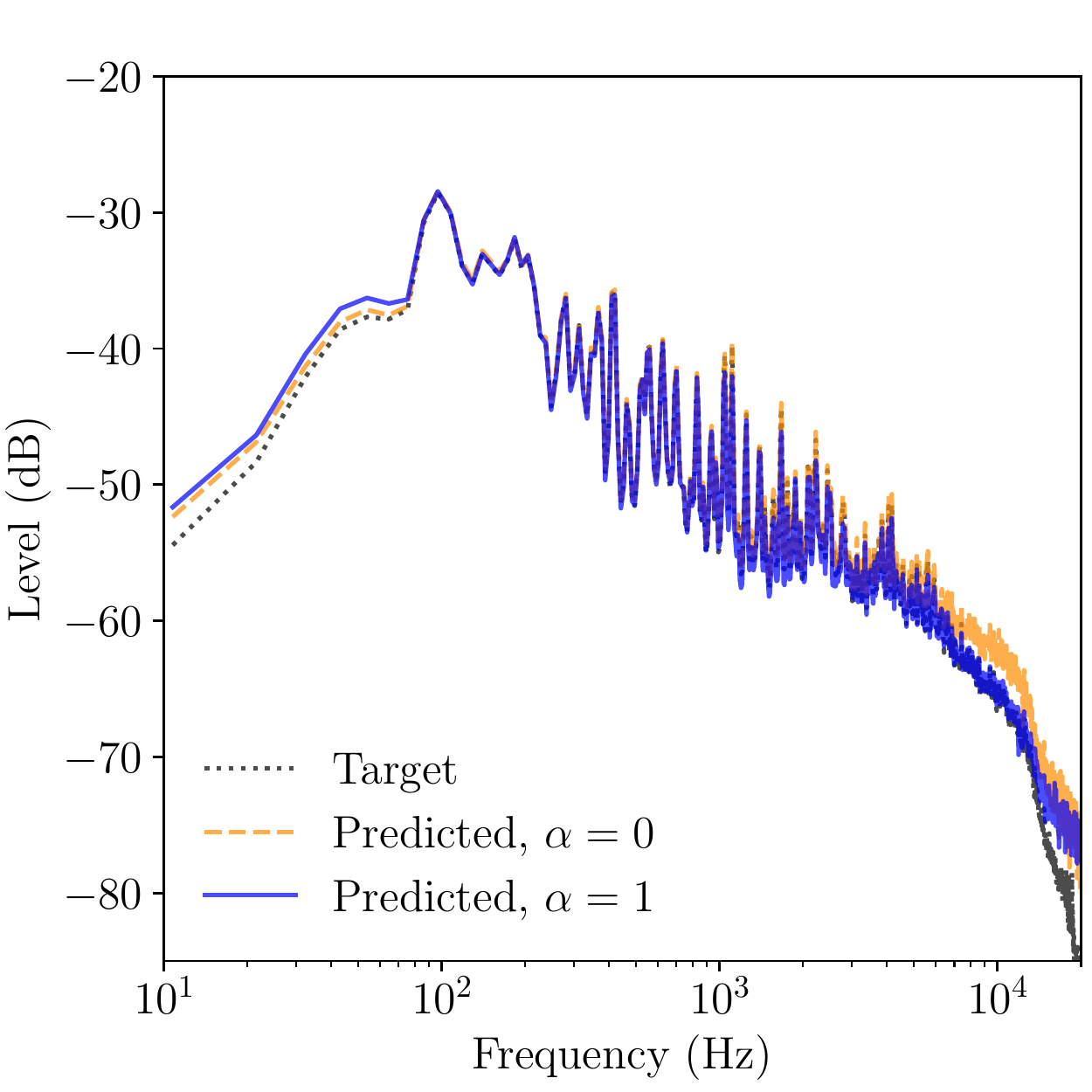}
\caption{\label{fig:spectra_freqL1}{
Power spectra for musical audio in the Test dataset\cite{suppl_materials} compressed with Comp-4C control parameters [-30, 2.5, .002, .03].  Here we see the effects of
weighting the L1 regularization in the loss
function Eq. (\ref{eq:loss}) exponentially by frequency ($\alpha=1$)
or not ($\alpha=0$):
weighting by frequency shifts a nontrivial amount of high frequency noise toward a proportionally small increase at
low frequencies.
Although noise is still clearly audible in both predicted outputs
(refer to Supplemental Materials
\cite{suppl_materials} to hear audio samples),
the result is that the listener perceives less overall noise in the output when the frequency-weighted L1 regularization is used.
}}
\vspace{-10pt}
\end{figure}

Although step responses are a useful diagnostic, the neural network model
approximates the input-output mappings it is trained on, and is ultimately intended for use
with musical sounds which may typically lack such sharp discontinuities.  Thus a comparison
of  compressor output for musical sounds is in order as well.  Figure \ref{fig:spectra_freqL1}
shows a comparison of frequency spectrum for a full-band recording
(\ie drums, bass, guitar, vocals) in the testing dataset.
It also shows that scaling the L1 regularization exponentially by frequency can yield
a reduction in high-frequency noise, sacrificing a proportionally smaller
amount of accuracy at low frequencies.

\vspace{-10pt}
\subsection{Analog Compressor: LA-2A}
\vspace{-10pt}
A primary interest in the application our method is not for cases in which a software plugin already exists, but rather for the profiling of analog units.
As an example, we choose the
Universal Audio's Teletronix LA-2A, an electro-optical compressor-limiter,\citep{la2a_lawrence1964} the controls for which consist of three  knobs and one switch.  Given that two of the knobs are only for input-output gain adjustment, for this study, we focus only on varying the ``Peak Reduction'' (PR) knob, and the ``Compress/Limit'' switch.  The switch is treated like any other knob, with limits chosen
arbitrarily to range from 0 for ``Compress" to 1 for ``Limit" (internally these are mapped to -0.5 and 0.5 to preserve zero-mean inputs to the neural network).  \footnote{Given the requirements of differentiability imposed by training via gradient descent optimization, one might expect the discontinuous nature of a switch to pose a problem for training, however we find this not to be the case.}  The dataset -- consisting of subsets for Training, Validation and Testing -- was created by assembling examples of music from sources with Creative Commons licenses, the authors' own recordings,
and synthetic waveforms such as those shown in Figure \ref{fig:synth_sounds}, all concatenated into a series of unique 15-minute WAV files at 44.1 kHz, and sent through the LA-2A at increments of 5 on the PR knob, for both  settings of the Comp/Lim switch.\footnote{We will make the dataset publicly available pending review.}

\begin{figure}[htp]
\includegraphics[width=0.8\columnwidth]{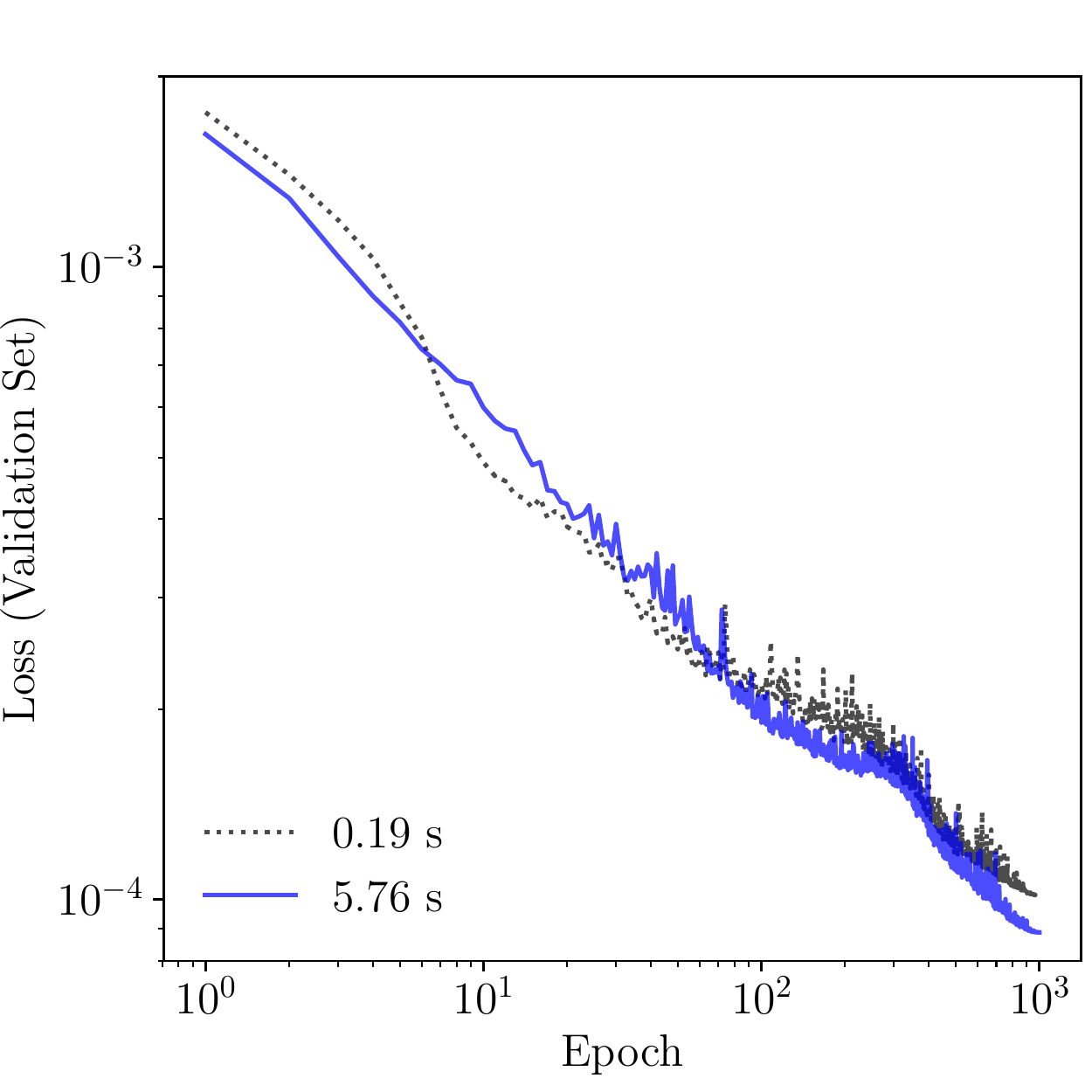}
\caption{\label{fig:la2a_vl}{Training history on the LA-2A  dataset for different lookback sizes. The ``kink'' near
epoch 300 is a common feature of the 1-cycle policy\citep{smith1,gugger}
when using an ``aggressive'' learning
rate (in this case, 7e-4).  Both runs achieve comparable losses
despite the longer lookback buffer needing nearly 5 times as much execution time.
 }}
 \vspace{-5pt}
\end{figure}

\begin{figure}[h!]
\includegraphics[width=\columnwidth]{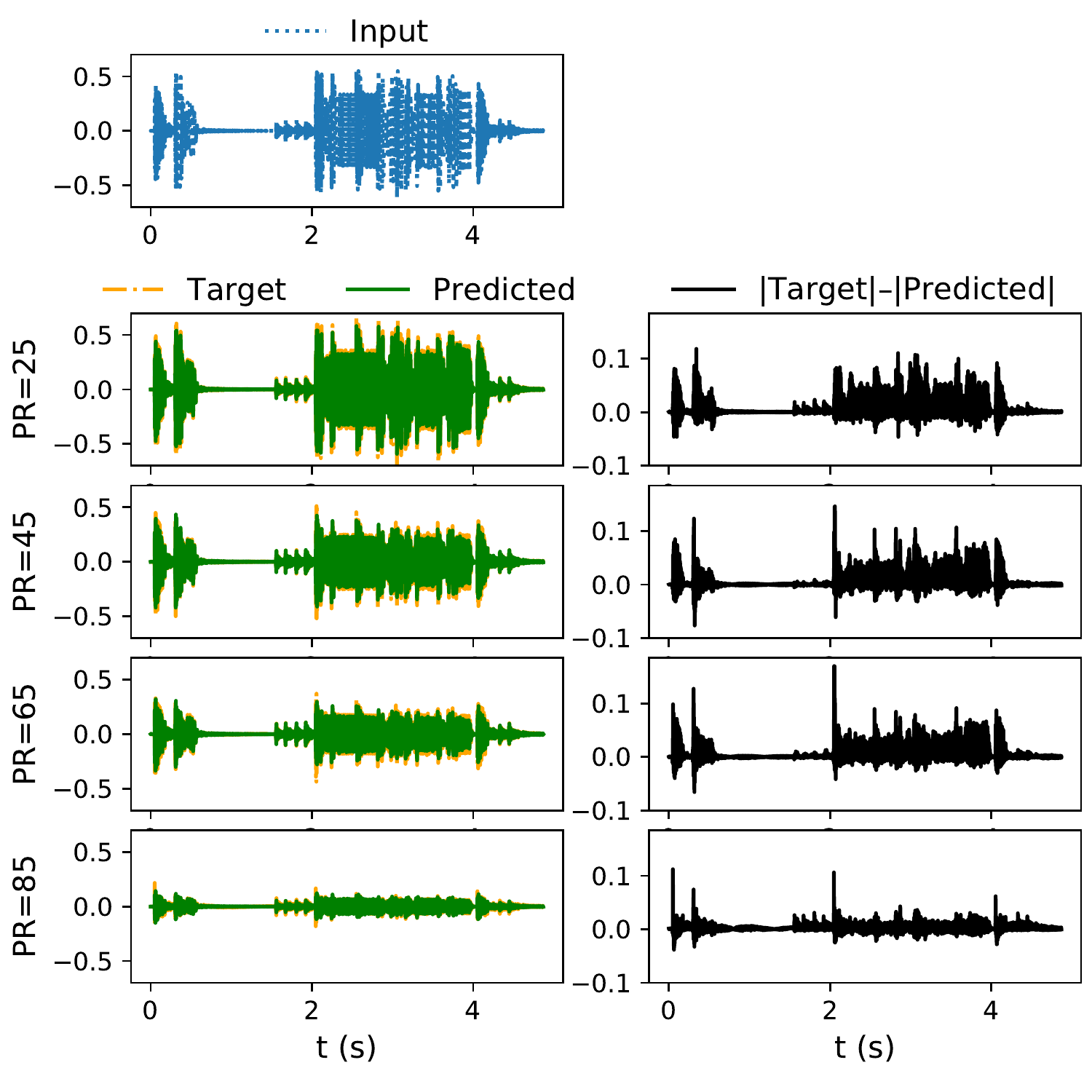}
 \vspace{-20pt}
\caption{\label{fig:la2a_collage}{Sample output for LA-2A using drum recordings
from the Testing dataset, for various values of the Peak Reduction control.
(The Comp/Lim switch setting had negligible effect on these outputs.)
We see that the model's predictions typically slightly
underestimate the target value for attack transients.
Audio samples are available in Supplemental Materials.\cite{suppl_materials}
}}
 \vspace{-10pt}
\end{figure}

Figure \ref{fig:la2a_vl} shows the loss on the validation set for the LA-2A for different lookback
sizes.  The dashed (black) line shows a model with an input size of 8192*2=16384 and took 15 hours to run, the solid (blue) line is a model with input size of 8192*27=221184 and took 72 hours.
Both runs used output sizes of 8192 samples.
In all our runs, a loss value of approximately 1e-4 is achieved,
regardless of the size of the model -- even for a lookback extending beyond
the ``5 seconds for complete release'' typically associated with the LA-2A.\citep{ua_catalog}
This indicates that the finite size of the lookback
window (or ``causality noise'') is not the primary source of error; this is consistent with the Comp-4C results (\eg see Figure \ref{fig:comp4c_train}). The primary source of error remains an ongoing subject of investigation.
Graphs of example audio waveforms from the Testing dataset are shown in Figure \ref{fig:la2a_collage}, where it is noteworthy that the
model will at times over-compress the onset of an attack as compared
to the true LA2A target response.

\section{\label{sec:concl}Conclusion}
In pursuit of the goal of capturing and modeling generic audio effects by means of artificial
neural networks, we have focused this study on dynamic range compressors
as a representative problem set because their nonlinear, time-dependent nature
makes them a challenging class of problems, and because they are a class of effects
of high interest in the field of musical audio production.  Rather than rely on domain-specific
knowledge of audio compressors in constructing our end-to-end system, our model learns
the effects the parameterized controls in the process of training on a large dataset
consisting of input-output audio pairs and the control settings used.

The results capture the qualities of the compressors sampled, although
the speed of execution and the residual noise in the neural network output suggest
that practical implementations of this method may await improvements in computer implementation and refinements to the model. We are interested in trying a model based on WaveNet \cite{van_den_oord_wavenet_2016,rethage_wavenet_2018}
or WaveRNN\citep{wavernn} for comparisons to our model regarding
speed and accuracy.

As the intent of this effort are the modeling of effects in general, more
work remains to probe the limits of our method toward a variety of other signal processing effects, both analog and digital, as well for the
construction of new effects by solving ``inverse problems'' such as
de-compression.\citep{gorlow2013decomp}

\begin{acknowledgments}
Scott H. Hawley wishes to thank Eric Tarr, William Hooper, and fast.ai (especially Jeremy Howard and Sylvain Gugger) for helpful discussions.
Stylianos Ioannis Mimilakis is supported by the by
the German Research Foundation (AB 675/2-1, MU 2686/11-1).
\end{acknowledgments}
\vspace{10pt}

\bibliography{signaltrain}

\end{document}